\documentclass[conference]{IEEEtran}
\IEEEoverridecommandlockouts
\usepackage{amsmath,amssymb,amsfonts}
\usepackage{graphicx}
\usepackage{textcomp}
\usepackage{xcolor}
\usepackage{url}
\usepackage{multirow}
\usepackage[square,numbers]{natbib}
\usepackage{tikz}

\usepackage[noend]{algpseudocode}
\usepackage{algorithm}

\newtheorem{definition}{Definition}
\newtheorem{example}{Example}

\usepackage[misc,geometry]{ifsym}

\def\BibTeX{{\rm B\kern-.05em{\sc i\kern-.025em b}\kern-.08em
    T\kern-.1667em\lower.7ex\hbox{E}\kern-.125emX}}
\DeclareMathOperator*{\argmax}{arg\,max}

\begin{document}

\title{Bound Propagation meets Constraint Simplification: Improving Logic-based XAI for Neural Networks}

\author{
\IEEEauthorblockN{
Ronaldo Gomes,
Jairo Ribeiro,
Luiz Queiroz,
Thiago Alves Rocha\IEEEauthorrefmark{1}\thanks{\IEEEauthorrefmark{1}The authors thank CNPq and CAPES for partial financial support.}
}
\IEEEauthorblockA{
\textit{Department of Computer Science, Instituto Federal do Ceará (IFCE)}, Maracanaú, Brazil
}%
\IEEEauthorblockA{
\{ronaldo.apoliano.gomes05, jairo.duarte.ribeiro01, luiz.fernando.paulino60\}@aluno.ifce.edu.br,
\IEEEauthorrefmark{1}thiago.alves@ifce.edu.br \Letter
}
}

\maketitle



\begin{abstract}
Logic-based methods for explaining neural network decisions offer formal guarantees of correctness and non-redundancy, but they often suffer from high computational costs, especially for large networks. In this work, we improve the efficiency of such methods by combining bound propagation with constraint simplification. These simplifications, derived from the propagation, tighten neuron bounds and eliminate unnecessary binary variables, making the explanation process more efficient. Our experiments suggest that combining these techniques reduces explanation time by up to 89.26\%, particularly for larger neural networks.


\end{abstract}



\section{Introduction}



Neural networks (NNs) have shown significant potential in tackling highly complex classification tasks~\cite{goodfellow16}. Despite their success, they are often regarded as opaque algorithms~\cite{koh2017understanding}. This means that the processes leading to their outputs from a given set of inputs are not inherently interpretable or explainable. In this context, for a given input and its corresponding output, an explanation is defined as a set of attributes that directly contribute to the final output. In other words, these attributes are sufficient to determine the result, whereas other attributes are not necessary. For example, given an input $\{sneeze=True$, $weight=70 \ kg$, $headache=True$, $age=40 \ years\}$ and its NN output \textit{flu}, a possible explanation could be $\{sneeze=True, headache=True\}$. That is, if an instance has the $sneeze=True$ and $headache=True$, the NN prediction is \textit{flu}, regardless of $weight$ and $age$ values.

Heuristic explainable AI (XAI) methods, such as LIME~\cite{ribeiro2016should} and Anchors~\cite{ribeiro2018anchors}, have been used to provide explanations for NNs and other machine learning (ML) models. However, these methods suffer from lack of correctness guarantees. Correctness guarantees are provided when there are no instances with the values specified in the explanation such that the NN makes a different prediction. Logic-based XAI approaches have been developed recently to offer explanations with such guarantees for a variety of ML models~\cite{shi2018, ignatiev2019abduction, izza2021explaining, gorji2022sufficient,wang2021probabilistic, bassan2023towards, audemard23computing, rocha2023logic}. While some recent explainability methods also offer formal guarantees, they typically focus on robustness or stability—e.g., ensuring that small input perturbations do not alter the explanation~\cite{lakkaraju2020robust, lin2023robustness}. Although important, these guarantees do not ensure correctness in the sense that the provided explanations are provably sufficient to reproduce the output of the model.



In contrast, logic-based approaches like the one proposed by Ignatiev et al.~\cite{ignatiev2019abduction} provide explanations that are both non-redundant and provably correct, which is the notion of guarantee we focus on in this work. This approach uses logical constraints—comprising Boolean combinations of linear (in)equalities involving binary and continuous variables—to compute non-redundant and provably correct explanations for NNs. A Mixed Integer Linear Programming (MILP) is then used to verify whether a subset of input attributes suffices to reproduce the output of the model. While this method ensures both correctness and succinctness of the explanation, it faces scalability challenges when applied to large neural networks.


In this work, we propose an approach to improve logic-based XAI for NNs by leveraging overapproximated bound propagation of neurons. This approach eliminates the need for the MILP solver in some cases, where we can determine that an input attribute is not necessary for the explanation. The overapproximation approach is sound but incomplete: if it determines that an input attribute is unnecessary, this is guaranteed to be correct; however, if it indicates that the attribute is necessary, there is still a possibility that the attribute is, in fact, not required. It is sound because the bounds produced by this method are guaranteed to include the actual values. However, it is incomplete because these bounds can be broader than the actual values. Then, in cases where it is not possible to determine that an input attribute is unnecessary for the explanation, the bounds obtained through propagation are used to simplify the logical constraints. Then, we use the MILP solver with the simplified constraints to determine whether the input attribute can be excluded from the explanation. With this, our goal is to develop a method that ensures correctness and provides redundancy-free explanations for NNs, while also enhancing scalability performance.

In our experiments, we compared explanation times between our approach and the original method presented by \citet{ignatiev2019abduction}. Results showed significant improvements, especially in larger NNs. For example, in one case, our approach achieved a reduction of over 89.00\%. These findings demonstrate the potential of combining a straightforward bound propagation approach with constraint simplification to enhance scalability in logic-based XAI, making explanations more efficient without compromising correctness.

In this work, we use a simple and coarse bound propagation method called Box. Existing literature~\cite{wang2018formal, gehr2018ai2, singh2019abstract} on NN verification has extensively explored Box and other bound propagation techniques for robustness analysis. While more advanced methods, such as Zonotope~\cite{gehr2018ai2} and DeepPoly~\cite{singh2019abstract}, offer greater precision in bound estimation, they are computationally more expensive. Exploring these more precise bound propagation methods will be left as future work.

Building on the strengths of approximated bound propagation, recent research in NN verification~\cite{wang2021beta, zhang2022general} has also focused on integrating these techniques with MILP solvers or direct branch-and-bound methods to enhance robustness certification. Typically, these approaches alternate between using relaxation techniques and MILP solvers to refine bounds at different stages of the verification process. In contrast, our approach to XAI employs a continuous application of bound propagation from start to finish. If this propagation does not conclusively determine that a given input attribute is unnecessary for the explanation, then we leverage the information obtained during the propagation process to simplify constraints before resorting to a MILP solver. In the other cases, the propagation alone is sufficient, eliminating the need to use the MILP solver entirely.



\section{Background}

This section aims to present basic concepts necessary for a general understanding of the work.

\subsection{First-order Logic over LRA}\label{subsec:logic}
In this work, we use first-order logic (FOL) to give explanations with guarantees of correctness. We use quantifier-free first-order formulas over the theory of linear real arithmetic (LRA). Then, first-order variables are allowed to take values from the real numbers $\mathbb{R}$. For details, see~\cite{kroening2016decision}. Therefore, we consider formulas as defined below:
\begin{equation}
        \begin{aligned}
             F, G &:= p \mid (F \wedge G) \mid (F \vee G) \mid (\neg F) \mid (F \to G),\\
             p &:= \sum^n_{i=1} w_i x_i \leq b, 
        \end{aligned}    
\end{equation}
such that $F$ and $G$ are quantifier-free first-order formulas over the theory of linear real arithmetic. Moreover, $p$ represents the atomic formulas such that $n \geq 1$, each $w_i$ and $b$ are fixed real numbers, and each $x_i$ is a first-order variable. Observe that we allow the use of other letters for variables instead of $x_i$, such as $s_i$, $z_i$, $q_i$. For example, $(2.5x_1 + 3.1x_2 \geq 6) \wedge (x_1=1 \vee x_1=2) \wedge (x_1=2 \to x_2 \leq 1.1)$ is a formula by this definition. Observe that we allow standard abbreviations as $\neg (2.5x_1 + 3.1x_2 < 6)$ for $2.5x_1 + 3.1x_2 \geq 6$.

Since we are assuming the semantics of formulas over the domain of real numbers, an \textit{assignment} $\mathcal{A}$ for a formula $F$ is a mapping from the first-order variables of $F$ to elements in the domain of real numbers. For instance, $\{x_1 \mapsto 2.3, x_2 \mapsto 1\}$ is an assignment for $(2.5x_1 + 3.1x_2 \geq 6) \wedge (x_1=1 \vee x_1=2) \wedge (x_1=2 \to x_2 \leq 1.1)$. An assignment $\mathcal{A}$ \textit{satisfies} a formula $F$ if $F$ is true under this assignment. For example, $\{x_1 \mapsto 2, x_2 \mapsto 1.05\}$ satisfies the formula in the above example, whereas $\{x_1 \mapsto 2.3, x_2 \mapsto 1\}$ does not satisfy it. Moreover, an assignment $\mathcal{A}$ \textit{satisfies} a set $\Gamma$ of formulas if all formulas in $\Gamma$ are true under $\mathcal{A}$.

A set of formulas $\Gamma$ is \textit{satisfiable} if there exists a satisfying assignment for $\Gamma$. To give an example, the set $\{(2.5x_1 + 3.1x_2 \geq 6), (x_1=1 \vee x_1=2), (x_1=2 \to x_2 \leq 1.1)\}$ is satisfiable since $\{x_1 \mapsto 2, x_2 \mapsto 1.05\}$ satisfies it. As another example, the set $\{(x_1 \geq 2), (x_1 < 1)\}$ is unsatisfiable since no assignment satisfies it. Given a set of formulas $\Gamma$ and a formula $G$, the notation $\Gamma \models G$ is used to denote \textit{logical consequence} or \textit{entailment}, i.e., each assignment that satisfies $\Gamma$ also satisfies $G$. As an illustrative example, let $\Gamma = \{x_1 = 2 , x_2 \geq 1\}$ and $G = (2.5x_1 + x_2 \geq 5) \wedge (x_1=1 \vee x_1=2)$. Then, $\Gamma \models G$. The essence of entailment lies in ensuring the correctness of the conclusion $G$ based on the set of premises $\Gamma$. In the context of computing explanations, as presented in~\cite{ignatiev2019abduction}, logical consequence serves as a fundamental tool for guaranteeing the correctness of predictions made by NNs.

The relationship between satisfiability and entailment is a fundamental aspect of logic. It is widely known that, for all sets of formulas $\Gamma$ and all formulas $G$, it holds that $\Gamma \models G$ iff $\Gamma \cup \{\neg G\}$ is unsatisfiable. For instance, $\{x_1 = 2, x_2 \geq 1),  \neg((2.5x_1 + x_2 \geq 5) \wedge (x_1=1 \vee x_1=2))\}$ has no satisfying assignment since an assignment that satisfies $\{x_1 = 2 , x_2 \geq 1\}$ also satisfies $(2.5x_1 + x_2 \geq 5) \wedge (x_1=1 \vee x_1=2)$ and, therefore, does not satisfy $\neg((2.5x_1 + x_2 \geq 5) \wedge (x_1=1 \vee x_1=2))$. Since our approach builds upon the concept of logical consequence, we can leverage this connection in the context of computing explanations for NNs.

\subsection{Classification Problems and Neural Networks}

In machine learning, classification problems are defined over a set of $n$ input attributes $\mathcal{F} = \{x_1, ..., x_n\}$ and a set of $k$ classes $\mathcal{K} = \{c_1, c_2,...,c_k\}$. In this work, we consider that each input attribute $x_i \in \mathcal{F}$ takes its values $v_i$ from the domain of real numbers. Moreover, each input attribute $x_i$ has an upper bound $ub_i$ and a lower bound $lb_i$ such that $lb_i \leq x_i \leq ub_i$, and its domain is the closed interval $[lb_i, ub_i]$. This is represented as a set of domain constraints or input space $\mathcal{D} = \{lb_1 \leq x_1 \leq ub_1,\text{ }lb_2 \leq x_2 \leq ub_2, ..., lb_n \leq x_n \leq ub_n \}$. For example, an attribute for the height of a person belongs to the real numbers and may have lower and upper bounds of $0.5$ and $2.1$ meters, respectively. Furthermore, $\mathcal{I} = \{x_1 = v_1, x_2 = v_2, ..., x_n = v_n\}$ represents a specific point or instance of the input space such that each $v_i$ is in the domain of $x_i$.

A NN is a function that maps elements in the input space into the set of classes $\mathcal{K}$. A feedforward NN is composed of $L+1$ layers of neurons. Each layer $l \in \{0, 1, ..., L\}$ is composed of $n_l$ neurons, numbered from $1$ to $n_l$. These layers and neurons define the architecture of the NN. Layer $0$ is fictitious and corresponds to the input of the NN, while the last layer $L$ corresponds to its outputs. Layers $1$ to $L-1$ are typically referred to as hidden layers. Let $x^l_i$ be the output of the $i$th neuron of the $l$th layer, with $i \in \{1,...,n_l\}$. The inputs to the NN can be represented as $x^0_i$ or simply $x_i$. Moreover, we represent the outputs as $x^L_i$ or simply $o_i$.

The values $x^l_i$ of the neurons in a given layer $l$ are computed through the output values $x^{l-1}_j$ of the previous layer, with $j \in \{1,...,n_{l-1}\}$. Each neuron applies a linear combination of the output of the neurons in the previous layer. Then, the neuron applies a nonlinear function, also known as an activation function. The output of the linear part is represented as $\sum_{j=1}^{n_{l-1}} w^{l}_{i,j} x^{l-1}_{j} + b^{l}_{i}$ where $w^{l}_{i,j}$ and $b^{l}_{i}$ denote the weights and bias, respectively, serving as parameters of the $i$th neuron of layer $l$. In this work, we consider only feedforward NNs with the Rectified Linear Unit ($\mathrm{ReLU}$) as activation function because it can be represented by linear constraints due to its piecewise-linear nature. This function is a widely used activation whose output is the maximum between its input value and zero. Then, $x^{l}_{i} = \mathrm{ReLU}(\sum_{j=1}^{n_{l-1}} w^{l}_{i,j} x^{l-1}_{j} + b^{l}_{i})$ is the output of the $\mathrm{ReLU}$. The last layer $L$ is composed of $n_L = k$ neurons, one for each class. Moreover, it is common to normalize the output layer using a Softmax layer. Consequently, these values represent the probabilities associated with each class. The class with the highest probability is chosen as the predicted class. However, we do not need to consider this normalization transformation as it does not change the maximum value of the last layer. Thus, the predicted class is $c_i \in \mathcal{K}$ such that $i = \argmax_{j \in \{1, ..., k\}} x^L_j$. Therefore, we represent a NN as a function $\mathcal{M}$ such that, for all instances $\mathcal{I}$, $\mathcal{M}(\mathcal{I}) = c \in \mathcal{K}$.






\subsection{MILP – Mixed Integer Linear Programming}

In Mixed Integer Linear Programming (MILP), the objective is to optimize a linear function subject to linear constraints, where some or all of the variables are required to be integers~\cite{milp1971}. MILP is a crucial technique in our work for determining the lower and upper bounds of each neuron in the NNs. For example, we utilize a minimization problem to determine the lower bound of neurons within NNs. This process involves formulating an objective function that seeks to minimize the lower bound, subject to constraints that reflect the behaviour of NNs. To illustrate the structure of a MILP, we provide an example below:
\begin{equation}
\label{eq:milp}
\begin{aligned}
\min \quad & y_1 \\
\textrm{s.t.} \quad & 1 \leq x_1 \leq 3\\
                    & 3x_1 - 2 \leq y_1 \\
                    & y_1 \leq 3x_1 - 2 - 0.5(1-z_1)\\
                    & 0 \leq y_1 \leq 8 z_1 \\
                    & z_1 \in \{0, 1\}
\end{aligned}
\end{equation}

In the MILP in (\ref{eq:milp}), we want to ﬁnd values for variables $x_1, y_1, z_1$ minimizing the value of the objective function $y_1$ among all values that satisfy the constraints. Variable $z_1$ is binary since $z_1 \in \{0, 1\}$ is a constraint in the MILP, while variables $x_1, y_1$ have the real numbers $\mathbb{R}$ as their domain.

An important observation is that a MILP problem without an objective function corresponds to a satisfiability problem, as discussed in Subsection~\ref{subsec:logic}. Given that the approach for computing explanations relies on logical consequence, and considering the connection between satisfiability and logical consequence, we employ a MILP solver to address explanation tasks. Additionally, throughout the construction of the MILP model, we may utilize optimization, specifically employing a MILP solver, to determine tight lower and upper bounds for the neurons of NNs. These tight bounds are computed immediately after training and prior to the explanation generation process. During the explanation phase, we rely solely on overapproximated bound propagation, which avoids the computational overhead of solving MILP problems in real-time.

\section{Logic-based XAI for NNs}

As previously mentioned, most methods for explaining ML models are heuristic, resulting in explanations that can not be fully trusted. Logic-based explainability provides results with guarantees of correctness and irredundancy. For NNs, \citet{ignatiev2019abduction} employed a logic-based approach based on linear constraints with binary and continuous variables. Given an input, this approach identifies a subset of input attributes sufficient to justify the correspondent output given by the NN. This method ensures correctness and minimality of explanations, which are referred to as \textit{abductive explanations}. An abductive explanation is a subset of attributes that form a rule. When applied, this rule guarantees the same model prediction. The formal definition below encapsulates this notion.

\begin{definition}[Abductive Explanation]
Let $\mathcal{I} = \{x_1 = v_1, ..., x_n = v_n\}$ be an input instance and $\mathcal{M}$ be a NN such that $\mathcal{M}(\mathcal{I}) = c \in \mathcal{K}$. An \emph{abductive explanation} $\mathcal{X}$ is a minimal subset of $\mathcal{I}$ such that for all $v_1' \in [l_1, u_1], ..., v_n' \in [l_n, u_n]$, if $v'_j = v_j$ for each $x_j = v_j \in \mathcal{X}$, then $\mathcal{M}(\{x_1 = v'_1, ..., x_n = v'_n\}) = c$.
\end{definition}

Minimality ensures that the explanation $\mathcal{X}$ does not include any redundant attributes. In other words, removing any attribute $x_j = v_j$ from $\mathcal{X}$ would result in a subset that no longer guarantees the same prediction $c$ over the defined bounds. It is important to note that a given instance may have multiple distinct abductive explanations. Different subsets of features may independently satisfy the conditions for an explanation while maintaining minimality. This means that there can be multiple ways to justify a prediction, each highlighting different aspects of the instance that are sufficient to ensure the same output.


The approach by \citet{ignatiev2019abduction} to obtain abductive explanations works as follows. First, the NN and the input space $\mathcal{D} = \{lb_1 \leq x_1 \leq ub_1,\text{ }lb_2 \leq x_2 \leq ub_2, ..., lb_n \leq x_n \leq ub_n \}$ are encoded as a set of formulas $\mathcal{F}$. Moreover, given an instance $\mathcal{I} = \{x_1 = v_1, x_2 = v_2, ..., x_n = v_n\}$ of the input space, the NN $\mathcal{M}$, and its corresponding prediction $\mathcal{M}(\mathcal{I}) = c_j \in \mathcal{K}$, we encode this fact as a formula $E$. This formula asserts that $o_j$ is the largest among all output neurons:
\[E = \bigwedge_{i=1, i \neq j}^{k} o_j > o_i.\]

Then, $\mathcal{I} \cup \mathcal{F}$ is satisfiable and $\mathcal{I} \cup \mathcal{F} \models E$. An abductive explanation $\mathcal{X}$ is calculated removing attribute by attribute from $\mathcal{I}$. For example, given $x_i=v_i \in \mathcal{I}$, if $(\mathcal{I} \setminus \{x_i=v_i\}) \cup \mathcal{F} \models E$, attribute $x_i$ is not necessary in an explanation and is removed. Otherwise, if $(\mathcal{I} \setminus \{x_i=v_i\}) \cup \mathcal{F} \not\models E$, then $x_i$ is kept since the same class cannot be guaranteed. This process is described in Algorithm \ref{algorithm1} and is performed for all attributes. Then, $\mathcal{X}$ is the result at the end of this procedure. This means that for the values of the attributes in $\mathcal{X}$, the NN makes the same classification, whatever the values of the remaining attributes.

\begin{algorithm}
\caption{Computing an Explanation $\mathcal{X}$} \label{algorithm1}
\begin{algorithmic}[1]
\Require{$\mathcal{F}$, $\mathcal{I}$, $E$}
\State $\mathcal{X} \gets \mathcal{I}$\;
\For{$x_i=v_i \in \mathcal{I}$}
\If{$(\mathcal{X} \setminus \{x_i=v_i\}) \cup \mathcal{F} \models E$}
\State $\mathcal{X} \gets \mathcal{X} \backslash \{x_i=v_i\}$\
\EndIf
\EndFor
\State \Return $\mathcal{X}$\;
\end{algorithmic}
\end{algorithm}


Since to check entailments $(\mathcal{X} \setminus \{x_i=v_i\}) \cup \mathcal{F} \models E$ is equivalent to test whether $(\mathcal{X} \setminus \{x_i=v_i\}) \cup \mathcal{F} \cup \{\neg E\}$ is unsatisfiable and $\mathcal{F}$, $\mathcal{X}$ and $\neg E$ are encoded as linear constraints with continuous and binary variables, such an entailment can be addressed by a MILP solver. In this work, we adopt the encoding introduced by \citet{tjeng2017evaluating} to represent a given NN as a set of formulas $\mathcal{F}$. Equations (\ref{eq:eq5.1}) to (\ref{eq:eq11}) describe this encoding, where $l= 1,\dots,L-1$ and, for each $l$, $j = 1,\dots,n_l$.
\begin{align}
    lb_i &\leq x_i \leq ub_i, \quad i=1,\dots,n \label{eq:eq10}\\
    x_j^l &\leq \sum_{i=1}^{n_{l-1}} w_{j,i}^l x_i^{l-1} + b_j^l - lb_j^l (1-z_j^l) \label{eq:eq5.1}   \\
    x_j^l &\geq \sum_{i=1}^{n_{l-1}} w_{j,i}^l x_i^{l-1} + b_j^l \label{eq:eq6} \\
    x_j^l &\leq ub_j^l z_j^l \label{eq:eq7} \\
    z_j^l &\in \{0,1\} \label{eq:eq8} \\
    x_j^l &\geq 0 \label{eq:eq9}\\
    o_j &= \sum_{i=1}^{n_{L-1}} w_{j,i}^L x_i^{L-1} + b_j^L, \quad j=1,\dots,n_L \label{eq:eq11}
\end{align}

In Equations (\ref{eq:eq5.1}) to (\ref{eq:eq11}), constants $ub^{l}_{j}$ represent the upper bound of the $j$th neuron of layer $l$. Similarly, constants $lb^{l}_{j}$ represent the lower bounds. The bounds $ub^{l}_{j}$ are defined by isolating variable $x^{l}_{j}$ from other constraints in subsequent layers. Then, $x^{l}_{j}$ is maximized to find its upper bound. A similar process is applied to find the lower bounds $lb^{l}_{j}$. This process can be performed using a MILP solver to obtain more precise bounds. This optimization is possible due to the bounds of the input attributes. Furthermore, these bounds can assist the solver in accelerating the computation of explanations. 

These bounds can be found tightly using MILP solvers and considering the entire domain of the input attributes, prior to the need for explanations. During the explanation process for a given instance, we use relaxed bounds according to our approach in this work. Since some input attributes are fixed during the process of finding explanations (i.e., $x_i = v_i$), if the overapproximated bound propagation finds tighter bounds, we replace the original bounds with those found during the explanation. These tighter bounds can help the MILP solver in cases where the propagation alone is unable to determine that the attribute is unnecessary. Once we have determined that an attribute is unnecessary, we revert to the original bounds before moving on to the next attribute.

\section{Our Approach}

This section describes the two approaches we combine to enhance the efficiency of computing logic-based explanations for NNs. First, we present the Box method for estimating neuron value ranges, emphasizing its role in identifying attributes that can be excluded from the explanation, which is central to our approach. Next, we discuss how the Box method refines neuron bounds and assists in simplifying MILP constraints, further improving the explanation process efficiency.

\subsection{Bound Propagation via Box for Logic-based XAI}

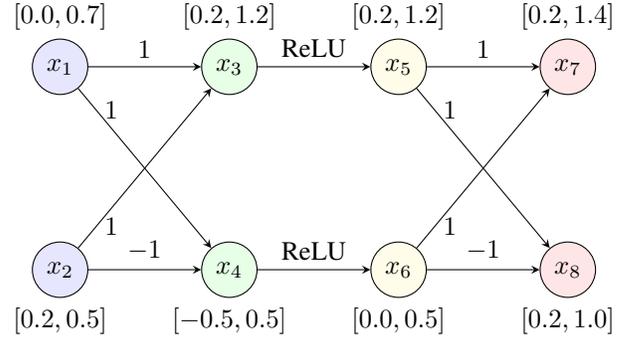
\begin{figure}[t]
\centering
\begin{tikzpicture}[x=2.5cm, y=1.5cm, >=stealth,scale=0.9]
\node[circle, draw, fill=blue!10, label=above:{$[0.0, 0.7]$}] (I1) at (0,1) {$x_1$};
\node[circle, draw, fill=blue!10, label=below:{$[0.2, 0.5]$}] (I2) at (0,-1) {$x_2$};

\node[circle, draw, fill=green!10, label=above:{$[0.2, 1.2]$}] (H1_pre) at (1,1) {$x_3$};
\node[circle, draw, fill=green!10, label=below:{$[-0.5, 0.5]$}] (H2_pre) at (1,-1) {$x_4$};

\node[circle, draw, fill=yellow!10, label=above:{$[0.2, 1.2]$}] (H1_post) at (2,1) {$x_5$};
\node[circle, draw, fill=yellow!10, label=below:{$[0.0, 0.5]$}] (H2_post) at (2,-1) {$x_6$};

\node[circle, draw, fill=red!10, label=above:{$[0.2, 1.4]$}] (O1) at (3,1) {$x_7$};
\node[circle, draw, fill=red!10, label=below:{$[0.2, 1.0]$}] (O2) at (3,-1) {$x_8$};

\draw[->] (I1) -- (H1_pre) node[midway, above] {$1$};
\draw[->] (I1) -- (H2_pre) node[pos=0.25, above] {$1$};
\draw[->] (I2) -- (H1_pre) node[pos=0.25, below] {$1$};
\draw[->] (I2) -- (H2_pre) node[midway, above] {$-1$};

\draw[->] (H1_pre) -- (H1_post) node[midway, above] {\text{ReLU}};
\draw[->] (H2_pre) -- (H2_post) node[midway, above] {\text{ReLU}};

\draw[->] (H1_post) -- (O1) node[midway, above] {$1$};
\draw[->] (H1_post) -- (O2) node[pos=0.25, above] {$1$};
\draw[->] (H2_post) -- (O1) node[pos=0.25, below] {$1$};
\draw[->] (H2_post) -- (O2) node[midway, above] {$-1$};

\end{tikzpicture}
\caption{Example of Neural Network}
\label{fig1}
\end{figure}


The Box method is a type of convex relaxation for functions that utilizes interval arithmetic to compute lower and upper bounds. Then, this method can be used to calculate the bounds of neurons~\cite{wang2018formal}. To illustrate the method, we consider the simple NN in Fig.~\ref{fig1}. This NN consists of an input layer, one hidden layer, and one output layer. Each layer consists of two neurons. There are two inputs to this NN, $x_1$ in the range $[0.0, 0.7]$ and $x_2$ in the range $[0.2, 0.5]$. To simplify the example, we separate each neuron in the hidden layer into two parts: one representing the output of the linear transformation and the other capturing the output of the ReLU activation. The weights for the linear transformations are represented by weights on the edges. The bias for each neuron is $0$ and has been omitted to simplify the example. Above or below each neuron, there is an interval representing the tight range of values the neuron can take. These initial intervals can be obtained through optimization. Note that if $x_5$ reaches the upper bound of $1.2$, then $x_1 = 0.7$ and $x_2=0.5$. However, under this input, the output at neuron $x_6$ becomes $0.2$. Therefore, the input $x_1 = 0.7$ and $x_2=0.5$ leads to the upper bound of neuron $x_7$. If the network had additional layers, the values of $x_7$ and $x_8$ would pass through the ReLU activation function, and their resulting values would propagate to the subsequent layers. 

The Box method can be used to estimate the range of values that each neuron in the NN can assume. However, as this method relies on overapproximation, the resulting intervals may be wider than the actual ranges of values. To demonstrate how the Box method works and to illustrate that it can produce looser intervals, we apply it starting from these input ranges \(x_1 = [0.0, 0.7]\) and \(x_2 = [0.2, 0.5]\). Initially, calculations are performed for the hidden layer neurons \(x_3\) and \(x_4\) as shown in (\ref{eq:eqx3}) and (\ref{eq:eqx4})-(\ref{eq:eq17}), respectively:
\begin{align}
    x_3 &= [0.0, 0.7] + [0.2, 0.5] = [0.2, 1.2] \label{eq:eqx3} \\
    x_4 &= [0.0, 0.7] + -1 \cdot [0.2, 0.5] \label{eq:eqx4}\\
        &= [0.0, 0.7] + [-0.5, -0.2] \\
        &= [-0.5 , 0.5] \label{eq:eq17}
\end{align}

\noindent Interval arithmetic is used to compute the ranges of values for $x_3$ and $x_4$. For $x_3$, this is done by adding the lower bounds and upper bounds of the intervals separately. For $x_4$, the first step is to multiply the interval $[0.2,0.5]$ by $-1$. This multiplication by a negative number reverses the interval, swapping the lower and upper bounds, and also inverting the signs of each bound. Next, the interval $[0.0,0.7]$ is added to the interval $[-0.5,-0.2]$, following the same procedure as for $x_3$: the lower bounds are added together, and the upper bounds are added together. Next, we apply the ReLU activation function to $x_3$ and $x_4$, respectively, which corresponds to neurons \(x_5\) and \(x_6\):
\begin{align}
    x_5 &= \text{ReLU}(x_3) = [\text{ReLU}(0.2), \text{ReLU}(1.2)]\\
        &= [0.2, 1.2] \\
    x_6 &= \text{ReLU}(x_4) = [\text{ReLU}(-0.5), \text{ReLU}(0.5)] \\
        &= [0.0, 0.5]
\end{align}

\noindent For $x_3=[0.2,1.2]$, both bounds are positive, so $x_5=[0.2,1.2]$. For $x_4=[-0.5,0.5]$, the negative lower bound is clipped to $0.0$, resulting in $x_6=[0.0,0.5]$. The ReLU function effectively sets negative values to $0.0$ while leaving positive values unchanged. Next, we can compute the output layer, which corresponds to neurons \(x_7\) and \(x_8\). The process for determining the bounds of $x_7$ and $x_8$ follows the same procedure used for $x_3$ and $x_4$. Then, $x_7 = [0.2, 1.7]$ and $x_8 = [-0.3 , 1.2]$. The intervals obtained through the Box method are wider than the actual possible value ranges for the neurons, as shown in Fig.~\ref{fig1}. For instance, for neuron $x_7$, the precise interval is $[0.2,1.4]$, while the Box method yields the broader interval $[0.2,1.7]$. In the context of the Box method, the following operations are applied to perform interval arithmetic, where \(lb_1, ub_1, lb_2, ub_2, p \in \mathbb{R}\) and \(p \geq 0\):
\begin{align}
    [lb_1, ub_1] + [lb_2, ub_2] &= [lb_1 + lb_2, ub_1 + ub_2] \\
    -1 \cdot [lb_1, ub_1] &= [-ub_1, -lb_1]  \\
    \text{ReLU}([lb_1, ub_1]) &= [\text{ReLU}(lb_1), \text{ReLU}(ub_1)]  \\
    p \cdot [lb_1, ub_1] &= [p \cdot lb_1, p \cdot ub_1] 
\end{align}

In our work, the goal of the Box method is to estimate the intervals during the explanation process. The procedure is similar to that of Algorithm~\ref{algorithm1}, where we remove attributes from $\mathcal{I}$ one by one. For each removed attribute, we compute the intervals for each neuron in the network using the Box method. The objective is to determine if the considered attribute is necessary for the explanation. The Box method can ensure that an attribute is not necessary when the lower bound of the neuron corresponding to the class $\mathcal{M}(\mathcal{I})$ is greater than the upper bounds of the other neurons in the output layer. In such a case, the MILP solver is not needed to perform the check $(\mathcal{X} \setminus \{x_i=v_i\}) \cup \mathcal{F} \models E$ in Algorithm~\ref{algorithm1}. Therefore, the objective of the Box method is to identify that an attribute is not necessary without resorting to a MILP solver.

For this purpose, we adapt Algorithm~\ref{algorithm1} to use the Box method at each iteration, prior to the check whether $(\mathcal{X} \setminus \{x_i=v_i\}) \cup \mathcal{F} \models E$. Additionally, we perform the propagation via the Box method based on the values in $\mathcal{X} \setminus \{x_i=v_i\}$ and the domain constraints $\mathcal{D}$. For attributes in 
$\mathcal{X} \setminus \{x_i=v_i\}$, we use the specific value of the instance $\mathcal{I}$. For the other attributes, we use the intervals according to $D$. To illustrate this idea, consider the following two examples.

\begin{example}
Let us revisit the NN in Fig.~\ref{fig1}. Consider the input values $x_1=0.7$ and $x_2=0.2$. Then, $x_7=1.4$ and $x_8=0.4$. Therefore, the NN predicts the class corresponding to the neuron $x_7$. According to our adaptation of Algorithm~\ref{algorithm1}, we first check if the Box method can determine whether the attribute $x_2$ is not necessary for the explanation. To do this, we start with the initial intervals $x_1=[0.7,0.7]$ and $x_2=[0.2,0.5]$. Using the Box method, we get $x_7=[1.1,1.7]$ and $x_8=[0.4,1.0]$. Since the lower bound of $x_7$ is greater than the upper bound of $x_8$, the Box method can guarantee that the attribute $x_2$ can be removed from the explanation. As a result, it was not necessary to use the solver to verify whether $(\mathcal{X} \setminus {x_i=v_i}) \cup \mathcal{F} \models E$. The Box method alone was sufficient to determine that the attribute $x_2$ is not required for the explanation, thus avoiding the computational overhead of invoking the solver.
\end{example}

\begin{example}\label{ex_box_fails}
Now, let us consider that Algorithm~\ref{algorithm1} begins by evaluating the attribute $x_1$. In other words, we will first check if the Box method can determine whether the attribute $x_1$ is not necessary for the explanation. To do this, we start with the initial intervals $x_1=[0.0,0.7]$ and $x_2=[0.2,0.2]$. Using the Box method, we get $x_3=[0.2,0.9]$ and $x_4=[-0.2,0.5]$. Then, $x_5=[0.2,0.9]$ and $x_6=[0.0,0.5]$. Finally, the Box method gives $x_7=[0.2,1.0]$ and $x_8=[-0.3,0.5]$. In this case, the lower bound of $x_7$ is smaller than the upper bound of $x_8$, meaning the Box method cannot guarantee that the attribute $x_1$ can be removed from the explanation. As a consequence, it becomes necessary to use the solver to determine whether the attribute $x_1$ can be removed from the explanation. This necessity arises because the intervals computed by the Box method are wider than the actual possible value ranges, leading to a lack of certainty in the decision solely based on Box intervals.
\end{example}

\subsection{Constraint Simplification from Bound Propagation}

The second part of our approach focuses on leveraging the bounds obtained from the Box method to assist the solver in performing satisfiability checks more efficiently. This is applied in cases where the Box method cannot guarantee that an attribute is unnecessary, requiring the use of a MILP solver. During the Box computation phase, we record the lower and upper bounds of all neurons computed by the method. During the construction of the MILP model, which can be done immediately after training the NN, we determine tight lower and upper bounds for its neurons. This step is performed prior to computing explanations. These tight bounds, computed during model construction, consider the domain constraints $\mathcal{D}$. When computing explanations, the Box method incorporates specific input values for attributes in $(\mathcal{X} \setminus \{x_i=v_i\})$. As a result, the Box method may produce bounds that are tighter than those computed earlier, since it uses specific input values for the input attributes rather than intervals. This allows the Box method to calculate narrower bounds in most cases, further refining the bounds.

For illustration, consider again Example~\ref{ex_box_fails}. Box can not guarantee that attribute $x_1$ can not be removed from the explanation. Then, we must use a MILP solver to check that. However, we can use the information found by Box. Consider the ranges from Fig.~\ref{fig1}. In the example, prior to the application of the Box method, the ranges of the neurons are $x_3 = [0.2, 1.2]$, $x_4 = [-0.5, 0.5]$, $x_5 = [0.2, 1.2]$, $x_6 = [0.0, 0.5]$, $x_7 = [0.2, 1.4]$, $x_8 = [0.2, 1.0]$. After applying the Box method, the bounds become narrower: $x_3 = [0.2, 0.9]$, $x_4 = [-0.2, 0.5]$, $x_5 = [0.2, 0.9]$, $x_6 = [0.0, 0.5]$, $x_7 = [0.2, 1.0]$, $x_8 = [-0.3, 0.5]$. Note that for neuron \( x_8 \), we can update its bounds to \( [0.2, 0.5] \) prior to invoking the solver. This is possible because the tight bounds, computed based solely on the domain constraints \( \mathcal{D} \), already ensure that the lower bound of \( x_8 \) is \( 0.2 \). The remaining bounds will retain the values determined by the Box method. Additionally, to illustrate how the MILP constraints change based on the new bounds identified, consider the constraint in Equation~\ref{eq:eq7} associated with neuron $x_5$. Initially, this constraint was $x_5 \leq 1.2 z_5$. However, with the updated upper bound obtained from the Box method, it can now be refined to $x_5 \leq 0.9 z_5$. Similarly, refinements can be applied to the constraints in Equation~\ref{eq:eq5.1} when a tighter lower bound is identified. These reductions in the neuron ranges, or equivalently, in the variables of the MILP model, can help the solver speed up the computation of explanations.

Therefore, leveraging the bounds obtained from the Box method, the simplification process compares the lower and upper bounds computed by Box with those initially determined. From this comparison, we retain the maximum of the lower bounds and the minimum of the upper bounds, effectively narrowing and refining the range to achieve tighter bounds. Additionally, if the lower bound found by the Box method for a neuron $x^l_j$ is greater than zero, we can simplify the constraints in Equations~(\ref{eq:eq5.1})-(\ref{eq:eq9}) to $x_j^l = \sum_{i=1}^{n_{l-1}} w_{j,i}^l x_i^{l-1} + b_j^l$. On the other hand, if the upper bound is less than or equal to zero, we can simplify these constraints to $x_j^l = 0$. In both cases, we are discarding a binary variable $z_j^l$. Since MILP solvers are generally more efficient with fewer binary variables, this reduction potentially improves computational efficiency. Following these simplifications, we solve a MILP problem to determine whether the attribute can be excluded from the explanation. After using the solver to determine whether the attribute should remain in the explanation, we discard the bounds obtained from the Box method. We then revert to the original bounds before proceeding to the next attribute. The Box method and simplification steps are performed independently for each attribute. This ensures that the computation is specifically tailored to the constraints and bounds associated with the current subset of $\mathcal{I}$ under evaluation.

\section{Experimental Results}




To evaluate the performance of our approach, we conducted experiments on five datasets from the UCI Machine Learning Repository\footnote{\url{https://archive.ics.uci.edu/ml/}}: Iris, Wine, Sonar, Digits, and MNIST. These datasets were selected to cover a range of complexities in terms of the number of attributes and classes. Iris is the simplest dataset, with 4 attributes and 3 classes, while MNIST is the most complex, containing 784 attributes and 10 classes.

For each dataset, we trained multiple NN architectures with varying depths and widths. The configurations tested include networks with 3 to 8 layers and 3 to 16 neurons per layer, depending on the dataset. The goal was to assess how explanation performance scales with increasing model complexity. We focus on small to medium-sized NNs due to the computational complexity involved in using MILP solvers. The NNs were implemented and trained using the TensorFlow library. The training process followed a standard 90\%-10\% data split, with 90\% of the data used for training and 10\% for testing. We used a \emph{batch size} of 8 and trained each NN for 100 \emph{epochs}.

\begin{table*}[t]
\caption{Comparison between INMS and our approach, highlighting total explanation times, solver execution times, and the impact of bound propagation and constraint simplification on neuron bounds and binary variable reduction.}
\label{tab:resultados_solver}
\begin{center}
\begin{tabular}{|c c|c|c|c|c|c|c|c|}
\hline
{\textbf{Dataset}} & {\textbf{NN}} & {\textbf{Exp (s)}} & {\textbf{Exp (s)}} & {\textbf{Solver (s)}} & {\textbf{Solver (s)}} & {\textbf{\% Bounds}} & {\textbf{\% Bin Vars}} & {\textbf{\% Bin Vars}} \\
{\textbf{  }} & {\textbf{  }} & {\textbf{INMS}} & {\textbf{Ours}} & {\textbf{INMS}} & {\textbf{Ours}} & {\textbf{tightened}} & {\textbf{Removed (Before)}} & {\textbf{Removed (Ours)}} \\
\hline
\multirow{2}{3em}{\textbf{Iris \\(4)}}
& 3L-3N & \textbf{0.92} & 1.15 & 0.24 & 0.05 & 92.83 & 0.00 & 31.61 \\ 
& 6L-6N & \textbf{1.28} & 1.79 & 0.56 & 0.05 & 96.04 & 21.21 & 56.43 \\
\hline
\multirow{2}{3em}{\textbf{Wine (13)}} 
& 3L-4N & \textbf{1.71} & 2.72 & 0.86 & 0.06 & 100.00 & 0.00 & 46.90 \\ 
& 6L-8N & \textbf{4.77} & 5.72 & 3.84 & 0.06 & 97.33 & 23.25 & 67.08 \\
\hline
\multirow{3}{3em}{\textbf{Sonar (60)}} 
& 3L-4N & \textbf{6.09} & 13.20 & 4.88 & 0.06 & 100.00 & 10.00 & 65.13 \\ 
& 6L-8N & \textbf{16.53} & 31.75 & 15.11 & 0.07 & 99.71 & 19.03 & 51.57 \\ 
& 8L-12N & 160.78 & \textbf{58.52} & 158.94 & 0.07 & 95.61 & 8.13 & 48.19 \\
\hline
\multirow{4}{3em}{\textbf{Digits (64)}} 
& 3L-8N & \textbf{186.55} & 189.13 & 174.20 & 0.58 & 100.00 & 0.00 & 52.13 \\ 
& 4L-12N & 623.03 & \textbf{287.15} & 609.17 & 0.56 & 100.00 & 0.00 & 64.67  \\ 
& 4L-16N & 1510.58 & \textbf{350.66} & 1494.95 & 0.56 & 100.00 & 0.00 & 68.44 \\
& 6L-12N & 3505.43 & \textbf{375.80} & 3489.62 & 0.58 & 99.61 & 5.71 & 66.65 \\ 
\hline
\multirow{3}{3em}{\scriptsize{\textbf{MNIST (784)}}} 
& 4L-8N & \textbf{1455.59} & 2443.89 & 1440.52 & 0.13 & 100.00 & 0.00 & 61.41 \\ 
& 4L-10N & 5267.60 & \textbf{2899.72} & 5249.62 & 0.13 & 100.00 & 0.00 & 62.59 \\
& 6L-8N & 9170.63 & \textbf{2513.24} & 9152.08 & 0.13 & 100.00 & 0.00 & 58.76 \\
\hline
\end{tabular}
\end{center}
\end{table*}

We compared two explanation methods: INMS (as proposed by \citet*{ignatiev2019abduction}) and our approach, which incorporates bound propagation and constraint simplification. Both methods produce the same explanations, ensuring correctness and minimality. The evaluation considered multiple performance metrics, including the total time to compute explanations across multiple instances, the total time spent by the MILP solver in generating explanations, the percentage of neuron bounds that were tightened, and the percentage of binary variables that were discarded. The percentage of neuron bounds that were tightened is computed as follows: For each instance and each attribute, we first apply the box method to determine whether the attribute can be safely removed from the explanation, as outlined in Algorithm~\ref{algorithm1}. If the box method fails, we then count the total number of neurons in the NN and the number of neurons whose bounds were tightened. Since different instances may require the MILP solver a varying number of times, neurons are counted multiple times, and each time we verify whether their bounds have been tightened. The final percentage is obtained by aggregating these counts across all instances. A similar process is used to compute the percentage of discarded binary variables. This metric is motivated by the fact that if the Box method determines a lower bound greater than zero or an upper bound less than or equal to zero, the constraints in Equations~(\ref{eq:eq5.1})-(\ref{eq:eq9}) can be simplified, effectively eliminating the binary variable $z_j^l$ in both cases. As MILP solvers generally perform better with fewer binary variables, this reduction can enhance computational efficiency.

Explanations were generated for instances randomly selected from the test data, ensuring consistency across both methods. The number of instances analyzed for each dataset was as follows: Iris (15), Wine (18), Sonar (21), Digits (180), and MNIST (20). Additionally, for the entailment checks in Algorithm~\ref{algorithm1}, we used IBM-CPLEX\footnote{\url{https://www.ibm.com/br-pt/analytics/cplex-optimizer}}, accessed via the DOcplex API, to model and handle MILP constraints. All details of the experiments can be seen in our repository\footnote{https://github.com/ronaldogomes96/explication-ann/tree/dev/ronaldo}. All experiments were executed on a MacBook Pro equipped with a 2.60 GHz Intel Core i7 processor, 16 GB of RAM, and running macOS 13.3. 

\subsection{Results}


TABLE I presents the data collected for each dataset. As shown in the columns \textbf{Exp (s) INMS} and \textbf{Exp (s) Ours}, there are a few cases where our method takes longer than INMS. This is likely due to the overhead introduced by the Box method, such as checking bounds and refining constraints, which can outweigh its potential benefits for simpler models. However, we observe that as NN complexity increases, by adding more layers or neurons, our approach significantly improves the total explanation time compared to the original method. One reason for this improvement is the efficiency of the Box method, which identifies unnecessary attributes in some cases, eliminating the need for the MILP solver and reducing computational costs. Another reason is the simplification process that refines the upper and lower bounds of neurons, often resulting in narrower bounds. This enables the solver to perform more efficiently.

In the Digits dataset, the simplest NN had a negligible time difference between INMS (186.55s) and our method (189.13s). However, in the most complex NN (6 layers, 12 neurons), our approach reduced the explanation time from 3505.43s to 375.80s, improving performance by over 89.26\%. In the MNIST dataset, with the most complex NN (6 layers, 8 neurons), our approach reduced the total explanation time from 9170.63s (INMS) to 2513.24s. This case involved the most complex NN for this dataset, which has the highest number of attributes in our experiment, highlighting the effectiveness of our approach.

Now, we turn to the results that help us better understand how and why our approach achieved superior performance. First, we focus on the total time spent by the MILP solver in generating explanations, as shown in the columns \textbf{Solver (s) INMS} and \textbf{Solver (s) Ours}. Notably, we observed significant improvements with our approach. For instance, for MNIST and the NN with 6 layers and 8 neurons per layer, the solver computation time with the original method took a total of 9152.08 seconds, while with our approach, it was reduced to just 0.13 seconds. While our method also involves additional steps—such as calculating bounds, and refining constraints—these tasks add minimal overhead compared to the substantial reduction in solver computation time. This highlights the clear advantage of our approach and justifies the overall performance gains. Regarding the percentage of neuron bounds that were tightened, our approach consistently tightened more than 90\%, and in some cases, even reached 100\%, as shown in the column \textbf{\% Bounds tightened}. This significant number of neuron bounds being tightened contributed to the improved solver scalability, as reflected in the substantial reduction in total explanation time.

Finally, we discuss the percentage of binary variables removed during the explanation process. To provide a more comprehensive comparison, we also included the original INMS method, which can simplify binary variables based on the input bounds before generating explanations. By doing so, we can assess whether our approach is capable of discarding even more binary variables throughout the explanation process. As shown in the columns \textbf{\% Bin Vars Removed (Before)} and \textbf{\% Bin Vars Removed (Ours)}, our method consistently removed more than 50\% of the binary variables across all datasets and NN configurations. Moreover, in every case, it increased the percentage of removed variables compared to the reduction achieved before the explanation process. These results highlight the effectiveness of our approach in further simplifying the MILP formulation, leading to improved solver efficiency and scalability.

\section{Conclusions}

In this work, we proposed an improved explanation method for NNs that integrates bound propagation and constraint simplification to enhance computational efficiency. Our approach was compared with the original method proposed by \citet{ignatiev2019abduction}, which introduced logic-based explanations for NNs and is a well-established approach in the literature. 


The experimental results demonstrate that the benefits of our method become more pronounced as the complexity of NNs increases. For example, in the Digits dataset, explanation time for the most complex NN dropped from 3505.43 seconds to just 375.80 seconds, and in the MNIST dataset, solver time was reduced from over 9150 seconds to 0.13 seconds also in the most complex NN. By consistently tightening more than 90\% of neuron bounds and removing over 50\% of binary variables, our method simplifies the explanation task at scale. These results highlight two important insights. First, our method is highly effective in identifying and eliminating irrelevant input attributes, which reduces the number of MILP queries. Second, the refinement of neuron bounds simplifies the MILP problems, leading to substantial solver speedups.




These findings suggest several promising directions for future work. One is to explore other method for propagating bounds, such as Zonotope and DeepPoly~\cite{gehr2018ai2, singh2019abstract}, to handle even larger NNs. Compared to the Box method, these approaches offer greater precision in the bounds they produce, but at the cost of higher computational complexity. Additionally, given the tight integration with MILP solvers, further improvements may come from customizing solvers to exploit the structure of the refined constraints generated by our method.




\bibliographystyle{IEEEtranN}

\bibliography{references}

\vspace{12pt}

\end{document}